\documentclass[aps,prl,twocolumn,showpacs,superscriptaddress,groupedaddress]{revtex4}  
\usepackage{graphicx}  
\usepackage{dcolumn}   
\usepackage{bm}        
\usepackage{float}
\usepackage{subfigure}

\newcommand{\ket}[1]{\ensuremath{\left| #1 \right \rangle}}

\begin{document}

\title{Detection of a single fundamental charge with nanoscale resolution in ambient conditions using the NV$^-$ center in diamond}

\author{Florian Dolde}
 \email{f.dolde@physik.uni-stuttgart.de}
 \affiliation{3. Physikalisches Institut, Research Center SCoPE and IQST, Universit\"at Stuttgart, Pfaffenwaldring 57. D-70550 Stuttgart, Germany}
\author{Marcus W. Doherty}
 \affiliation{Laser Physics Centre, Research School of Physics and Engineering, Australian National University, Australian Capital Territory 0200, Australia}
 \author{Julia Michl}
  \affiliation{3. Physikalisches Institut, Research Center SCoPE and IQST, Universit\"at Stuttgart, Pfaffenwaldring 57. D-70550 Stuttgart, Germany}
\author{Ingmar Jakobi}
 \affiliation{3. Physikalisches Institut, Research Center SCoPE and IQST, Universit\"at Stuttgart, Pfaffenwaldring 57. D-70550 Stuttgart, Germany}
\author{Boris Naydenov}
 \affiliation{Institut f\"ur Quantenoptik and IQST, Universit\"at Ulm, Ulm D-89073, Germany}
\author{Sebastien Pezzagna}
 \affiliation{Physikalisches Institut, Universit\"at Leipzig, 04103 Leipzig, Germany}
\author{Jan Meijer}
 \affiliation{Physikalisches Institut, Universit\"at Leipzig, 04103 Leipzig, Germany}
\author{Philipp Neumann}
 \affiliation{3. Physikalisches Institut, Research Center SCoPE and IQST, Universit\"at Stuttgart, Pfaffenwaldring 57. D-70550 Stuttgart, Germany}
\author{Fedor Jelezko}
 \affiliation{Institut f\"ur Quantenoptik and IQST, Universit\"at Ulm, Ulm D-89073, Germany}
\author{Neil B. Manson}
 \affiliation{Laser Physics Centre, Research School of Physics and Engineering, Australian National University, Australian Capital Territory 0200, Australia}
\author{J\"org Wrachtrup}
 \affiliation{3. Physikalisches Institut, Research Center SCoPE and IQST, Universit\"at Stuttgart, Pfaffenwaldring 57. D-70550 Stuttgart, Germany}

\begin{abstract}
Single charge detection with nanoscale spatial resolution in ambient conditions is a current frontier in metrology that has diverse interdisciplinary applications. Here, such single charge detection is demonstrated using two nitrogen-vacancy (NV) centers in diamond. One NV center is employed as a sensitive electrometer to detect the change in electric field created by the displacement of a single electron resulting from the optical switching of the other NV center between its neutral (NV$^0$) and negative (NV$^-$) charge states. As a consequence, our measurements also provide direct insight into the charge dynamics inside the material.

\end{abstract}

\pacs{76.30.Mi,61.72.jn,76.70.Hb,84.37.+q,07.50.Ls}
\maketitle

Single charge detectors with nanoscale spatial resolution that operate under ambient conditions have diverse interdisciplinary applications as probes of physical phenomena \cite{set1,set2,afm1}, components of quantum and nano-devices \cite{set3,qdot}, and as high-performance sensors of chemical and biological species \cite{chemsensor1,chemsensor2}.
The detection of  elementary charges is a long-standing endeavour, with a suite of low temperature/pressure techniques available, including single-electron transistors \cite{set1,set2,set3}, scanning probe microscopy \cite{afm1,afm2,afm3} , electric field-sensitive atomic force microscopy \cite{afm4}, electromechanical resonators \cite{graphene,nanoresonator} and nanowire field-effect transistors \cite{nwire}.
Yet, few techniques are available that operate under both ambient temperature and pressure and can detect small numbers of elementary charges \cite{mems}.
None of which currently achieve nanoscale resolution.
Recently, the atomic-sized negatively-charged nitrogen-vacancy (NV$^-$) center in diamond was demonstrated in ambient conditions to be a high-sensitivity electrometer with potential nanoscale resolution \cite{efield}.
Indeed, it was projected that the NV$^-$ center could be used to detect the electric field of a single electron at a distance of $\sim150$ nm within one second of averaging.
Here, we demonstrate a vital first advance in single charge detection using the NV$^-$ center by sensing the presence of a single electron at a distance of 25 $\pm 2$ nm.
\\
\indent
Beyond electrometry, the NV$^-$ center has a range of impressive applications including, high-sensitivity nano-magnetometry \cite{waldherr12,staudacher13,mamin13} and -thermometry \cite{toyli13,neumann13,kucsko13}, quantum information science \cite{dolde13,bernien13}, and bioimaging \cite{biomag}.
Each of these applications exploit some combination of the center's remarkable properties under ambient and extreme conditions \cite{toyli12,doherty13}: strong fluorescence that enables the detection of atom-sized single centers \cite{gruber97}, long-lived ground state electron spin coherence and optical spin-polarization/readout \cite{review}.
More specifically, the NV center is a $C_{3v}$ point defect in diamond consisting of a substitutional nitrogen - carbon vacancy pair
orientated along $\left\langle 111 \right\rangle$ crystal axis.
The negative charge state (NV$^-$) is formed from the neutral NV$^0$ when the center traps an excess electron.
NV$^-$ is characterized by an optical ZPL (zero phonon line) at $\sim1.945$ eV (637 nm) that is accompanied by phonon sidebands that extend to higher/lower energy in absorption/emission \cite{review}.
Additionally, NV$^-$ exhibits a ground $^3A_2$ spin triplet level with a spin-spin splitting between the $m_s=0$ and $\pm1$ spin sub-levels of $D\sim2.87$ GHz at room temperature.
The spin state of the ground triplet level can be prepared and read out via optical excitation to the excited $^3E$ triplet level.
Spin-orbit and spin-spin mixing of the triplet levels makes the ground state spin resonances susceptible to electric and crystal strain fields \cite{efield,doherty12,vanoort90}.
The long-lived ground state spin coherence, thus enables small electric field shifts of the spin resonances to be sensitively detected using optically detected magnetic resonance (ODMR) techniques.
\begin{figure}[hbtp]
\begin{center}
\mbox{
\includegraphics[width=1\columnwidth] {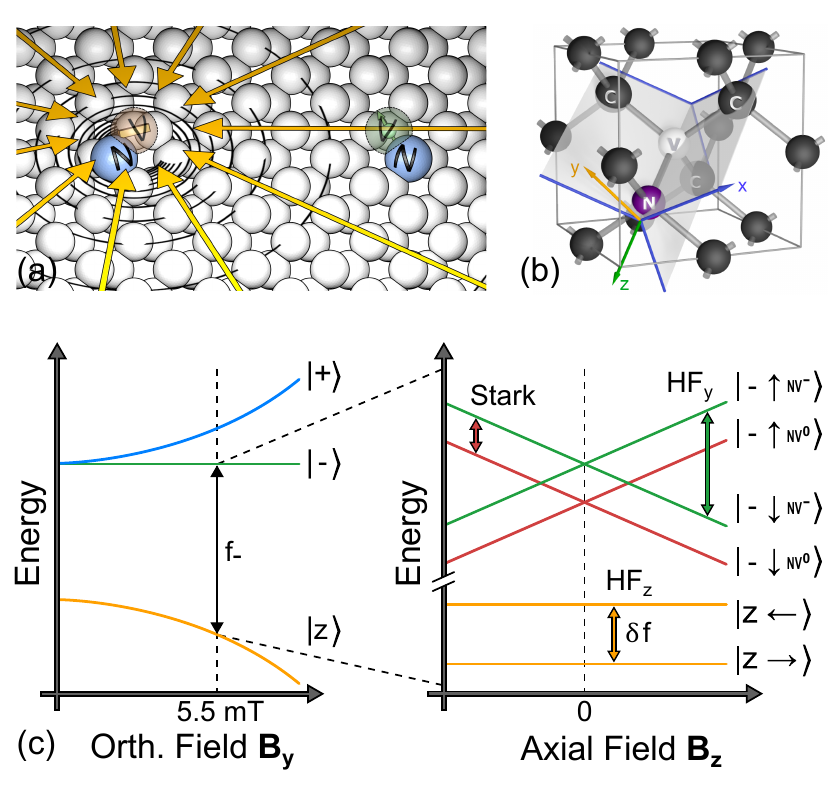}}

\caption{(color online)
(a) Schematic of the NV centers under investigation. On the right NV~A is depicted as the electrometer NV, and on the left NV~B is responsible for the electric field. 
(b) diamond unit cell containing an NV center.
Mirror planes highlight the trigonal symmetry.
The Cartesian coordinate axes $x,y$ and $z$ (blue, orange, green) are shown.
The $x$-axis can point along any of the three displayed mirror planes.
The vacancy, the nitrogen and the nearest neighbor carbon atoms are marked with V, N and C respectively.
(c) Left: Electron spin levels affected by a transverse magnetic field $B_{\perp}$ (here: $B_y$)
of varying strength.
Right: Effects on energy levels of hyperfine (HF) coupled electron spin nuclear spin pair due to additionally applied electric and longitudinal magnetic fields (Stark and $B_z$ respectively).
See text for discussion. 
}
\label{fig:fig1}
\end{center}
\end{figure}
\\
\indent
ODMR electrometry may be further introduced via the NV$^-$ ground state spin-Hamiltonian \cite{doherty12,supmat}
\begin{eqnarray}
H&=&(D+ k_{\parallel} E_z ) (S_z^2-2/3)+\frac{g_e\mu_B}{h} \vec{S}\cdot\vec{B}\nonumber \\
&&-k_{\perp}E_x (S_x^2-S_y^2)+k_\perp E_y (S_x S_y+S_y S_x )
\label{eq:spinhamiltonian}
\end{eqnarray}
where $\vec{S}$ are the $S=1$ dimensionless electron spin operators, $\mu_B$ is the Bohr magneton, $g_e\sim2.003$ is the electron g-factor \cite{loubser78,felton09},
$h$ is the Planck constant, $\vec{B}$ and $\vec{E}$ are the magnetic and electric fields, respectively, $k_\parallel=0.035(2)$ kHz m/V and $k_\perp=1.7(3)$ kHz m/V are the electric susceptibility parameters \cite{vanoort90}, and the spin coordinate system is defined such that the $z$ coordinate axis coincides with the center's trigonal symmetry axis and the $x$ axis is contained in one of the center's mirror planes (see figure \ref{fig:fig1}b%
).
Given the orders of magnitude difference between $k_\parallel$ and $k_\perp$, the electron spin is most sensitive to electric fields that are transverse to the center's trigonal axis (i.e. within the $x-y$ plane).
Sensitivity is enhanced by tailoring the electron spin eigenbasis $\{\ket{z},\ket{-},\ket{+}\}$ using a transverse magnetic field, such that the $\ket{0}\leftrightarrow\ket{\pm}$ electron spin resonance frequencies
$f_\pm$
are linearly susceptible to both axial and transverse electric field components \cite{efield,doherty12,supmat}
\begin{eqnarray}
f_\pm \approx f_\pm(0) + k_\parallel E_z\mp k_\perp E_\perp\cos(2\phi_B+\phi_E)
\label{eq:resfreq}
\end{eqnarray}
where $\tan \phi_B=B_y/B_x$, $\tan \phi_E=E_y/E_x$, $B_\perp=\sqrt{B_x^2+B_y^2}$, $E_\perp=\sqrt{E_x^2+E_y^2}$, and
$f_\pm(0)$
are the resonance frequencies in the absence of an electric field, which depend on $B_\perp$, but not $\phi_B$.
The bare electric field shift is given by $\Delta f_\pm = f_\pm - f_\pm(0)$.
As demonstrated in Ref. \onlinecite{efield}, the effects of the transverse orientations of the electric and magnetic fields on the spin resonances are coupled by the final term  in equation \ref{eq:resfreq}, which can be observed by rotating the magnetic field in the transverse plane.
\\
\indent
The neutral charge state NV$^0$ is characterized by an optical ZPL at $\sim 2.156$ eV (575 nm) accompanied by phonon sidebands. 
In the absence of light, the equilibrium NV charge state is determined by the local distribution of electron donors and acceptors \cite{review}. Alternatively, the equilibrium NV charge state may be controlled via gate voltages applied to the diamond \cite{hauf11,grotz12}.
Different photoconversion processes enable controlled optical switching of the NV charge state \cite{rittweger09,waldherr11,beha12,han12,aslam13}.
Under red (637-575 nm) excitation, NV$^-$ is selectively excited and subsequently an electron is transferred to the conduction band, converting the center to NV$^0$.
Likewise, under blue ($<$490 nm) excitation, NV$^-$ is directly ionized, converting the center to NV$^0$.
Under green (490-575 nm) excitation, both charge states are excited and photoconversion occurs in both directions.
However, the negative charge state is the dominant one with respect to occurrence and fluorescence intensity in the latter spectral range.
The rate of each photoconversion process depends quadratically on the excitation intensity.
\\
\indent
Improved understanding of the charge dynamics of the NV center \cite{loubser78,manson05,positron,mita96,gali09,weber10,ranjbar11,rittweger09,waldherr11,beha12,han12,aslam13} is particularly important to the performance of NV$^-$ in its various applications.
However, there has not yet been a direct observation of the excess electron whose presence/absence determines the center's charge state.
Here, we perform such a direct observation that unequivocally confirms the current charge state assignments and also provides insight into the microscopic behavior of the excess electron.
\begin{figure}[hbtp]
\begin{center}
\mbox{

\includegraphics[width=1\columnwidth] {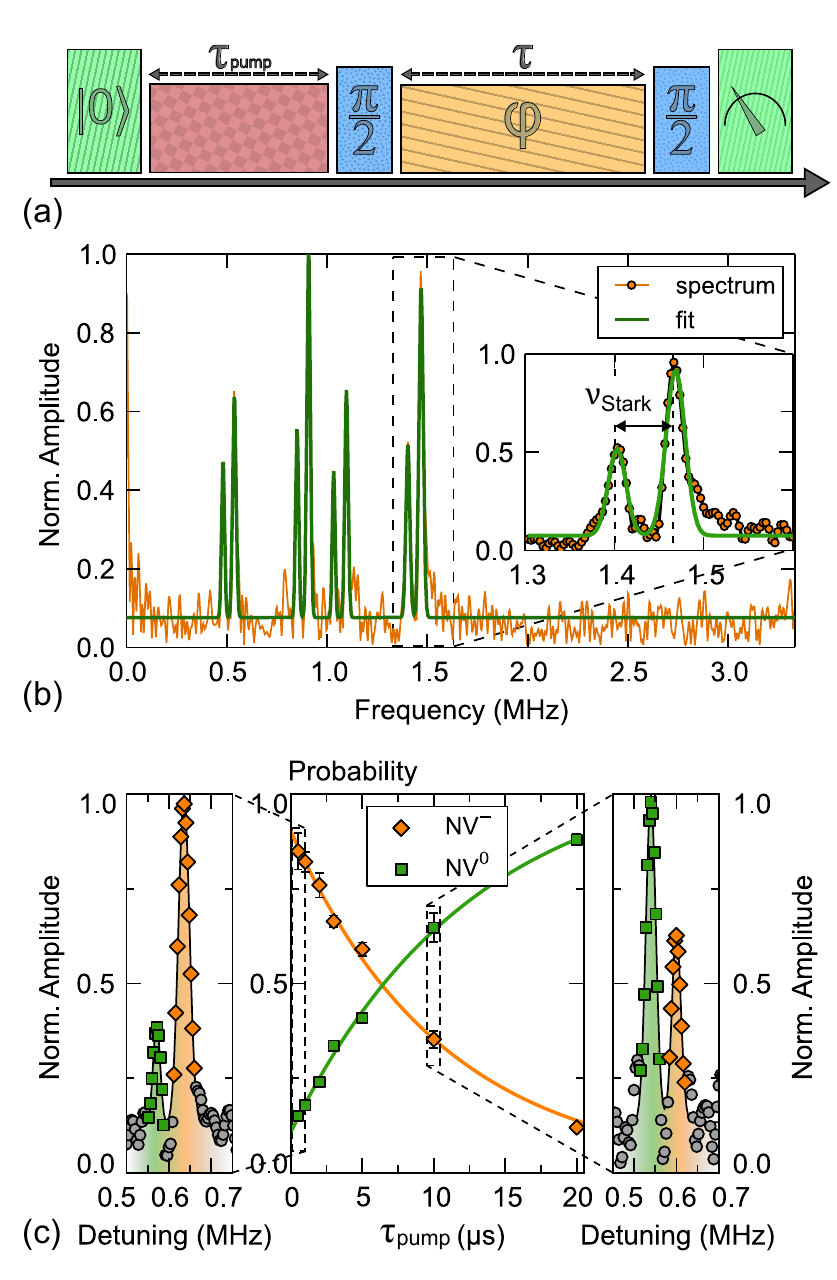}
}
\caption{(color online)
(a) Ramsey measurement sequence including charge state preparation of NV~B (red laser pulse), spin state initialization and readout of NV~A (green laser + fluorescence detection), microwave pulses for spin control (blue) and free evolution of the spin superposition state (yellow; accumulated phase $\varphi$ during time $\tau$).
(b) Example Ramsey spectrum revealing four pairs of resonance lines (orange: measurement data, green: multiple Gauss fit).
The pairs are split apart from each other due to hyperfine interaction (see text).
Inset: The splitting within each pair results from the Stark effect.
(c) The relative intensity (center) of the Stark-split lines of NV~A (left and right inset) can be influenced by pumping NV~B from its negative into its neutral charge state with pumping duration $\tau_{\mathrm{pump}}$.
Thus demonstrating that the Stark effect originates from different charge states of NV~B. }
\label{fig:fig2}
\end{center}
\end{figure}
\\
\indent
In our experiments, we employed a pair of implanted $^{15}$NV$^-$ centers that were oriented along different $\left[ 111 \right]$
directions in type IIa diamond and whose positions have been previously established using super-resolution microscopy \cite{dolde13,pezzagna11}.
One center (NV~A) was employed as an electrometer to detect the change in electric field created by the displacement of the excess electron of the other center (NV~B) when it is optically switched from NV$^-$ to NV$^0$.
By selectively addressing the ODMR transitions of NV~A, we were able to perform electrometry on NV~A using polarized green ($532\,$nm) spin-polarization and -readout laser pulses (timed with fluorescence detection) and optically switch the charge state of NV~B using polarized red ($638\,$nm) pump laser pulses \cite{supmat}.
The length of the red pulse $\tau_\mathrm{pump}$ controlled the probability that NV~B was in a given charge state during the ODMR measurement \cite{supmat}.
Similar to Ref. \onlinecite{efield}, the ODMR electrometry utilized a static transverse bias magnetic field with magnitude $B_{\perp}\approx 5.5\,$mT,
a smaller auxiliary magnetic field $\delta \vec{B}$ that tuned the net magnetic field (see fig.~\ref{fig:fig3}a),
and a Ramsey-type microwave pulse sequence with free spin evolution time $\tau$.
The transverse bias magnetic field split the $\ket{z}\leftrightarrow\ket{\pm}$ electron spin resonances and the microwave pulses were tuned to the lower frequency $\ket{z}\leftrightarrow\ket{-}$ electron spin resonance (see figure \ref{fig:fig1}).
The full measurement scheme is summarized in figure \ref{fig:fig2}.
\\
\indent
The Fast Fourier Transform (FFT) of the electron spin Ramsey oscillation of NV~A reveals the spectrum (e.g. figure \ref{fig:fig2}) of the $\ket{z}\leftrightarrow\ket{-}$ electron spin transition.
The spectrum contains four pairs of spectral lines, where one member of the pair has lower intensity than the other.
Each pair corresponds to a single hyperfine resonance of the magnetic hyperfine interaction between the electron spin and the $^{15}$N nuclear spin (refer to figure \ref{fig:fig1}), which is described by the addition of the potential \cite{review}
\begin{eqnarray}
V_\mathrm{hf}=A_\parallel S_zI_z+A_\perp(S_xI_x+S_yI_y)
\end{eqnarray}
to the spin-Hamiltonian (\ref{eq:spinhamiltonian}), where $\vec{I}$ are the $I=1/2$ dimensionless nuclear spin operators, $A_\parallel=3.03(3)$ MHz and $A_\perp=3.65(3)$ MHz are the $^{15}$N hyperfine parameters \cite{felton09}.
To second-order, the $m_I=\pm1/2$ nuclear spin projections are degenerate for the $\ket{\pm}$ electron spin states,
but are mixed for $\ket{z}$ and split by $\delta f\approx2A_\perp B_\perp/D$ (see fig.~\ref{fig:fig1}b).
Consequently, each electron spin resonance becomes two hyperfine lines that are split by $\delta f$.
If a small axial magnetic field $B_z$ is present, then the nuclear spin projections split for the $\ket{\pm}$ electron spin states and each electron spin resonance now becomes a four-line hyperfine structure (see fig.~\ref{fig:fig2}b).
\\
\indent
The presence of pairs of lines with different intensities in the Ramsey spectrum may be understood by first noting that the spectrum represents a statistical average of the spin resonances over the measurement ensemble \cite{doherty12}.
Since during some of the measurements NV~B was NV$^-$ and during others it was NV$^0$, the Ramsey spectrum contains two sets of hyperfine lines that correspond to the two charge states of NV~B.
The two sets of lines are shifted with respect to each other due to the electric field shift at NV~A that results from the change in charge at NV~B.
Since the electric field shift is smaller than the hyperfine splittings, the two sets of lines appear to form pairs of lines with different intensities.
The integrated intensities of the two sets of lines are directly related to the the probabilities that NV~B was NV$^-$ or NV$^0$ during a measurement.
To confirm our interpretation, we varied the illumination time $\tau_\mathrm{pump}$ with the red charge state switching laser in order to vary the charge state probabilities of NV~B.
Figure \ref{fig:fig2} clearly demonstrates that the ratio of the two sets of lines follow inversely related single exponential curves with $\tau_\mathrm{pump}$, which agrees well with the expected variation of the charge state probabilities provided by the current model of the NV$^-$$\rightarrow$NV$^0$ photoconversion process \cite{waldherr11,supmat}.
\begin{figure}[hbtp]
\begin{center}
\mbox{
\includegraphics[width=1\columnwidth] {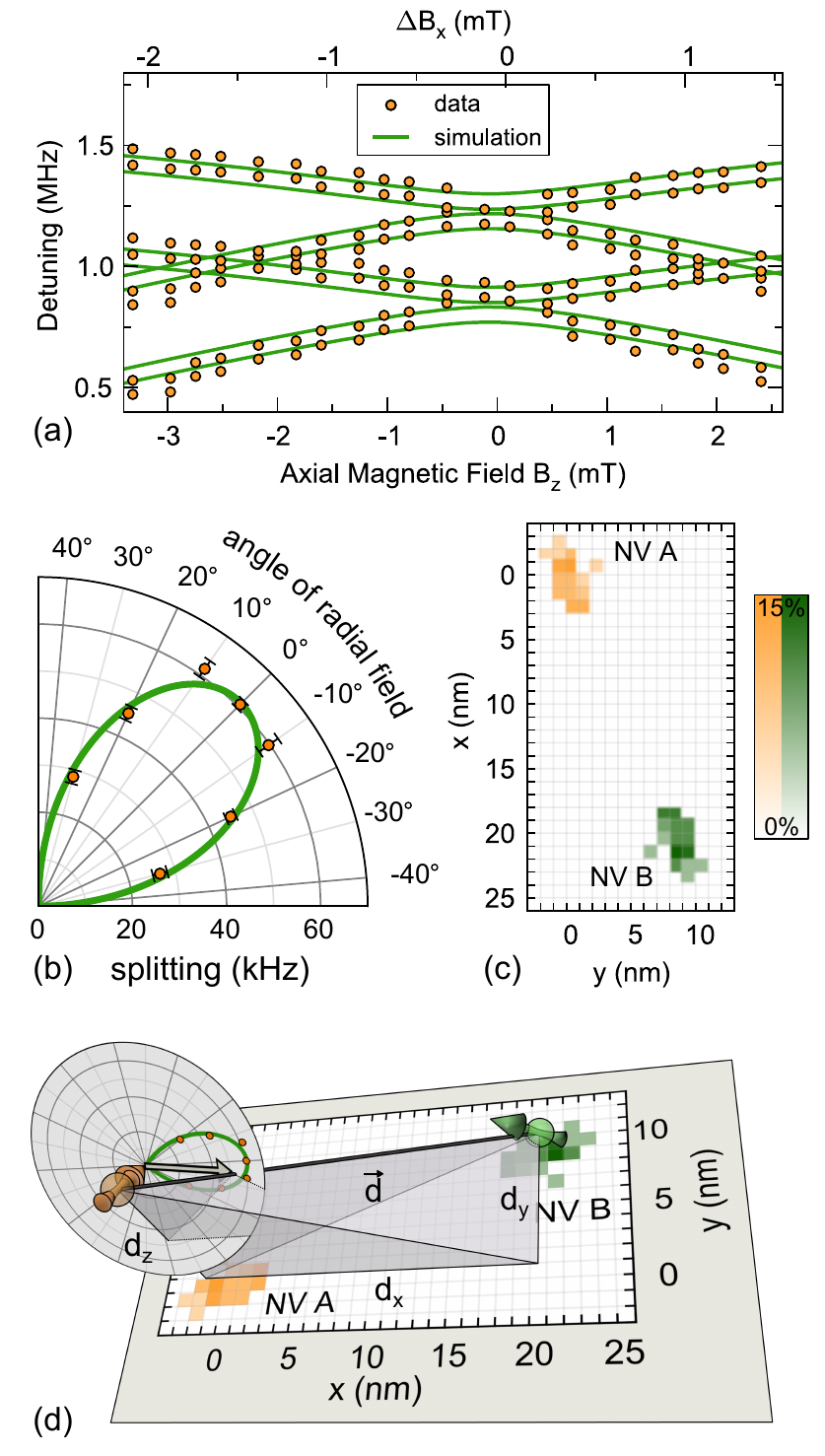}
}
\caption{(color online)
(a) Electron spin resonance frequencies as obtained from Ramsey oscillations
for different auxiliary magnetic fields ($\delta \vec{B} = \Delta B_x\hat{x} + B_z\hat{z}$)
(b) Polar plot of the electric field shift of a single hyperfine line as a the transverse magnetic field $B_{\perp}$ is rotated.
(c) Super-resolution microscopy image of NV~A and NV~B.
(d) Combination of (b), (c) and the known orientations and positions of the centers.}
\label{fig:fig3}
\end{center}
\end{figure}
\\
\indent
Given the non-trivial interplay of the magnetic, electric and hyperfine interactions that govern the observed spin resonances, in order to precisely measure the electric field shift, we recorded the Ramsey spectra for different magnetic field configurations and fit the spectral line frequencies using numerical solutions of the complete spin-Hamiltonian $H+V_\mathrm{hf}$.
Figure \ref{fig:fig3}(a) depicts the results of a sweep of the auxiliary magnetic field $\delta \vec{B}$,
which yields $\Delta f_-=66\pm7\,$kHz.

Of note for NV$^-$ electrometry, the electric field shifts are observed over a much larger range of $B_z$ than in the first electrometry demonstration \cite{efield}, which is due to the larger transverse bias field employed in this work.
This outcome promises that NV$^-$ electrometry may be successfully implemented in the future with less sophisticated magnetic field alignment. 
\\
\indent
Figure \ref{fig:fig3}(b) depicts the variation of the electric field shift of a single hyperfine line as the transverse magnetic field $B_{\perp}$ was rotated around the $z$-axis.
The polar pattern of figure \ref{fig:fig3}(b) displays one of the `leaves' of the `four-leaf' pattern predicted by the $\cos(2\phi_B+\phi_E)$ angular dependence of the spin resonances (\ref{eq:resfreq}) and observed in Ref.~\onlinecite{efield}.
Figure \ref{fig:fig3}(d) combines the ODMR electrometry results with the known positions and orientations of the centers to demonstrate that the polar pattern is orientated towards NV~B from NV~A.
Given the expected angular dependence, this suggests that the transverse electric field at NV~A is similarly orientated.
Figure \ref{fig:fig3}(d) also demonstrates that the displacement vector $\vec{d}$ connecting NV~A and NV~B.  Electric field simulation yields a field angle of $6\pm 4^\circ$ and and $\Delta f_{sim}=70 \pm 10\,$kHz
Considering the orientation of an electric field at NV~A generated by a charge at NV~B, this angle and the observed net electric field shift of
$\Delta f_-\approx 66\,$kHz
implies that the shift due to the axial component of the electric field is $k_\parallel E_z < 0.3\,$kHz, which is too small to be detected.
Specifying $k_\parallel E_z \sim 0$, the fit of the magnetic field data of figure \ref{fig:fig3}(a) yields the transverse electric field shift $k_\perp E_\perp=66\pm7$ kHz and a field angle of $0.5\pm4^\circ$ which is in good agreement with the calculated values form the distance vector.
\\
\indent
Accounting for the relative permittivity of diamond ($\epsilon_r=5.7$), the measured transverse electric field is that of a single electron located at a transverse distance of $25\pm2$ nm from NV~A, which is consistent with the super-resolution microscopy measurement of $27\pm3$ nm. Noting that the measured electric field is the difference in the electric field at NV~A due to the change of charge at NV~B, this result may be interpreted as the displacement of the excess electron at NV~B when it is NV$^-$ to a position beyond the range of detection ($>40$ nm).
\\
\indent
Combining our evidence obtained from optically controlling the NV charge state, observing the OMDR as a function of magnetic field and interpreting super-resolution microscopy, it is clear that we successfully employed one NV center as an electrometer to detect the single excess electron that determines the charge state of another NV center located $\sim25$ nm away. Thus demonstrating single charge detection using NV$^-$ electrometry under ambient conditions and unequivocally confirming the current charge state assignments. Furthermore, our measurements provided direct insight in the microscopic behavior of the excess electron.
\begin{acknowledgements}
This work was supported by the ARC (DP120102232), SFB~TR/21, SFB~716, Forschergruppe 1493 as well as EU projects SIQS and ERC SQUTEC and the Max Planck Society.
\end{acknowledgements}

\end{document}


\title{Supplementary material to `Detection of a single fundamental charge with nanoscale resolution in ambient conditions using the NV$^-$ center in diamond'.}

\author{Florian Dolde}
\author{Marcus W. Doherty}
\author{Julia Michl}
\author{Ingmar Jakobi}
\author{Boris Naydenov}
\author{Sebastien Pezzagna}
\author{Jan Meijer}
\author{Philipp Neumann}
\author{Fedor Jelezko}
\author{Neil B. Manson}
\author{J\"org Wrachtrup}

\maketitle

\section{Experimental details}

\subsection{Creation of two proximal NV centers}
Given the sensitivity of NV- electrometry,  to detect the change in charge at one NV center by a second NV center, the two centers must be <150 nm apart.
The chances of finding two NV centers with this separation in the dilute samples necessary for single NV addressing are remote. 
Therefore we used a high energy mask implantation. As mask we chose nanochannels in mica fabricated by GeV heavy ion bombardment and successive etching of the ion tracks \cite{pezzagna11}. 
An implantation energy of 1 MeV per nitrogen ion was chosen to have an average implantation depth of about 750 nm, thereby shielding the NV centers from charge traps and charge noise on the surface without having a too greater straggle.
A NV pair with a distance of $ 25\pm 2 $ nm was selected.
The relative positions of the centers have been previously established through interpretation of superresolution measurements and the magnetic dipolar coupling of the electron spins of the two centers \cite{dolde13}.

\subsection{Measurement setup}
We used a home build confocal setup to address the NV centers with diffraction limited resolution.
Under optical saturation, our free space confocal microscope collects $\approx$ 300 kcounts/s per single NV center for a $\left\langle 110\right\rangle$ surface close to saturation.
The NV center's ground state was controlled by microwaves applied with a lithographically fabricated coplanar wave guide on the diamond surface.
A set of coils around the setup for the precise control of the magnetic field amplitude (up to $\approx$6.5 mT) and orientation.. 
The green laser (532 nm, modulated by an AOM) and the red laser (638 nm, modulated by internal electronics) were combined by a beam splitter and coupled into a photonic crystal fiber to achieve the same beam pathway.

\subsection{Polarization cross section of a NV center}
\begin{figure}[htb]
	\centering
		\includegraphics{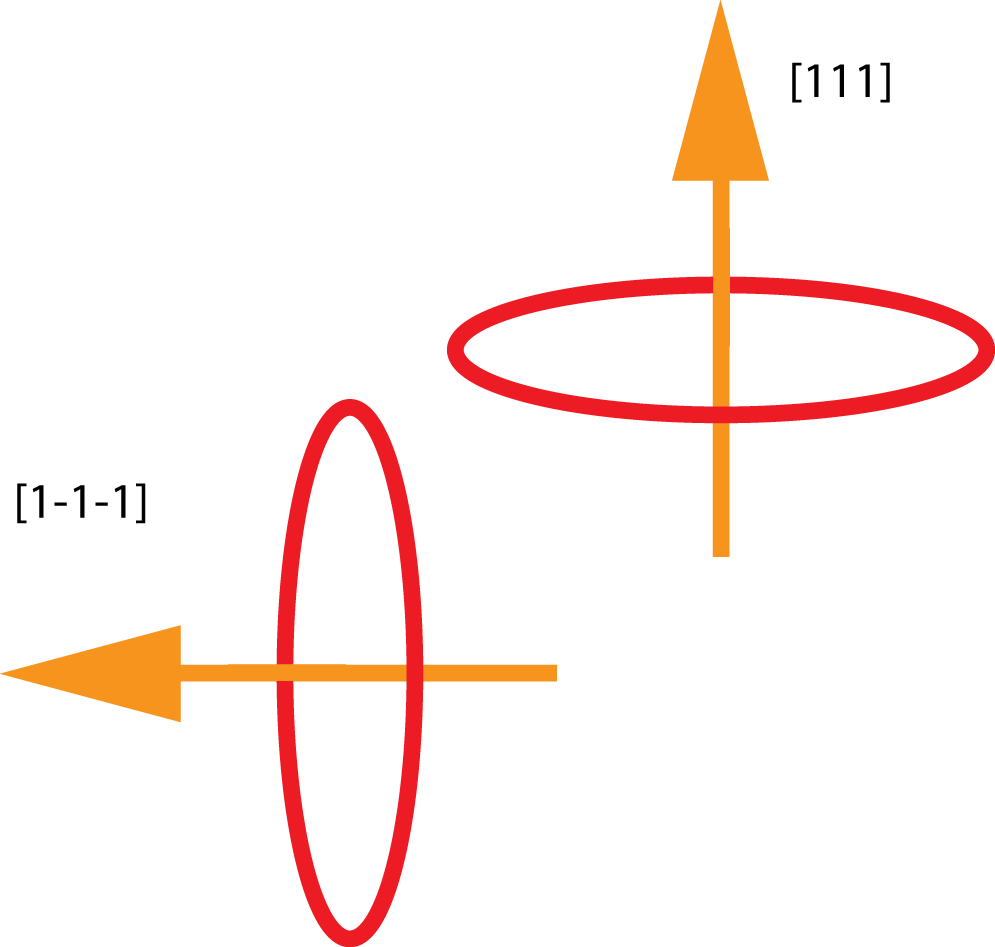}
	\caption{Absorption cross section for different NV orientations for a [100] diamond surface}
	\label{fig:supplementary_figure_absorprtion_crosssection}
\end{figure}

The NV center has two transition dipoles orthogonal to the NV axis. This leads to an circular absorption cross section of linear polarized light along the NV axis. Since the used sample has a $\left\langle 110 \right\rangle$ surface  the NV centers are not orthogonal to the surface, but are tilted by half the tetrahedral angle. 
Therefore the circular absorption cross section becomes an elliptical projection on the diamond surface. 
Since NV A and NV B have different orientations, their projection on the surface is orthogonal, such that linear polarized light has a minimal absorption cross section on one NV if it is maximal on the other NV. 
The  absorption cross section is given by
\[\sigma(\theta,\alpha)=\left(1-\sin\theta\sin\alpha\right)\sigma_{\text{max}}\]
where $\theta$ ($\approx 55^\circ$) is the angle between the NV axis and the surface normal and $\alpha$ is the angle between the light polarization and projection of the NV axis on the surface.
Since photoionization is believed to be a two photon process \cite{aslam13} the ionization ratio for red light aligned with one NV center is significantly higher than than for the nonaligned case. 
This allows for asymmetric charge state pumping of the NV centers thereby allowing the ODMR measurement of NV A to be relatively unperturbed.

\section{Theoretical details}

\subsection{Model of the NV$^-$$\rightarrow$NV$^0$ photoconversion process}

\begin{figure}[hbtp]
\begin{center}
\mbox{
\subfigure[]{
\includegraphics[width=0.7\columnwidth] {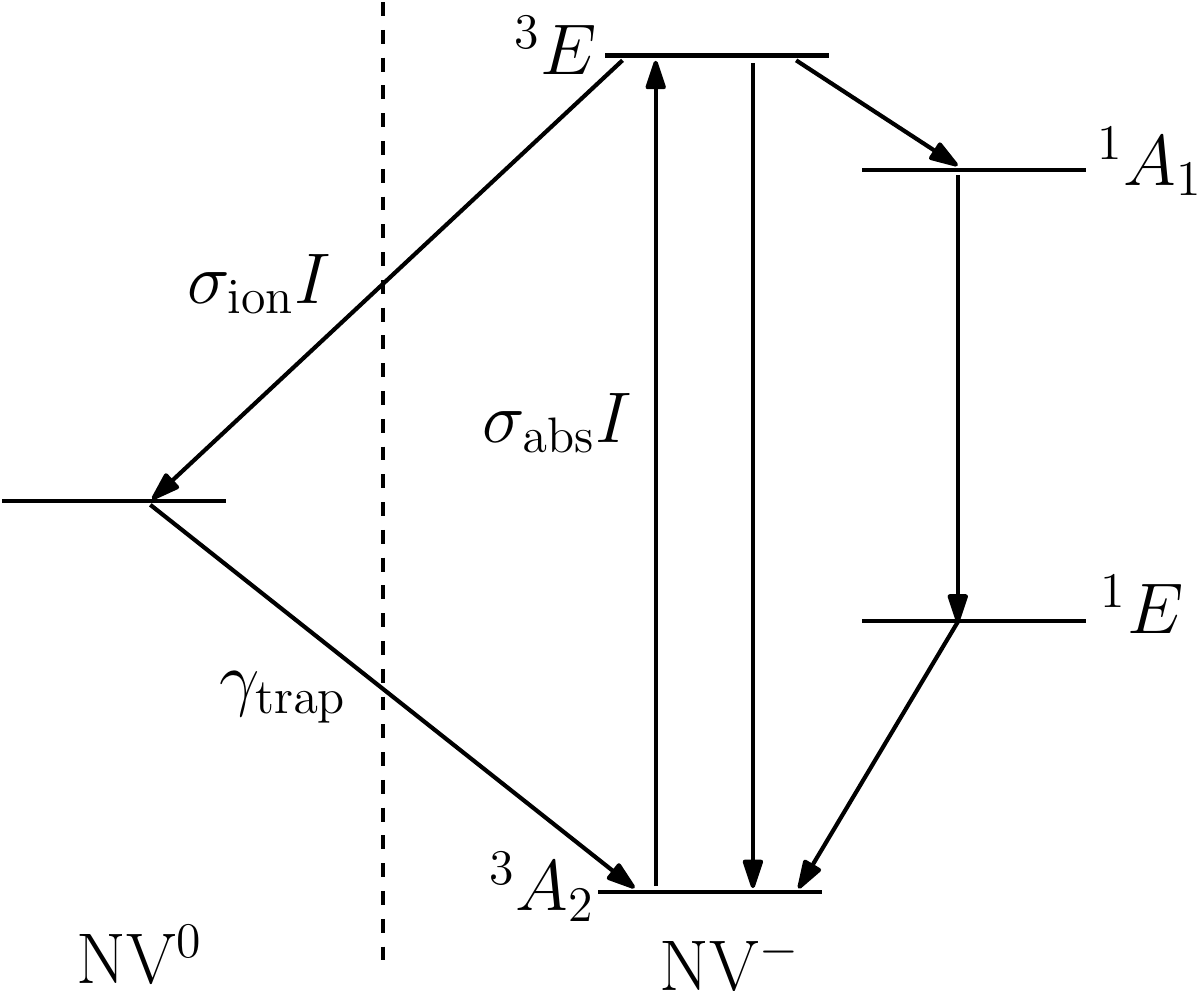}}
}
\mbox{
\subfigure[]{
\includegraphics[width=0.6\columnwidth] {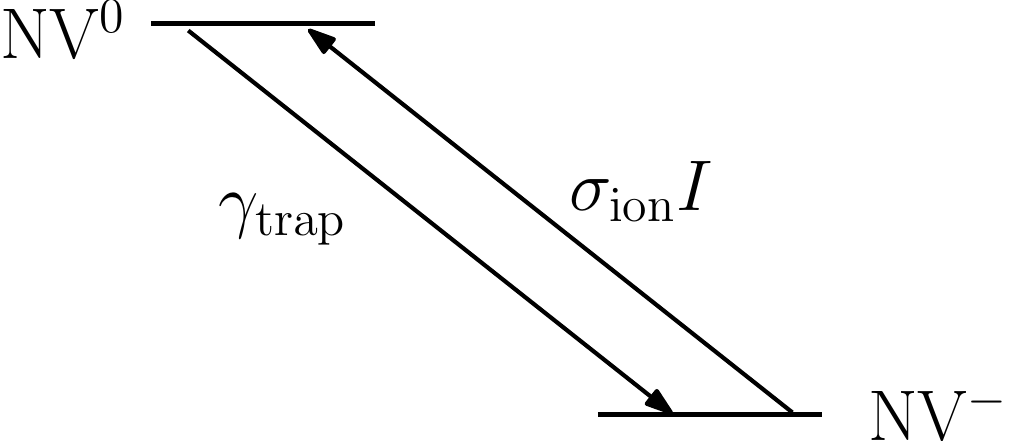}}
}
\caption{(a) A five-level rate equation model of the optical cycle of NV$^-$ including photoconversion to NV$^0$, which is represented by a single level on the left hand side. (b) The reduced two-level model corresponding to the saturation of the NV$^-$ optical transition.}
\label{fig:photoconversion}
\end{center}
\end{figure}

Under red (637-575 nm) excitation, NV$^-$ is selectively excited. A five-level rate equation model of the optical cycle of NV$^-$ is depicted in figure \ref{fig:photoconversion}(a).\cite{aslam13} Absorption of one photon transfers NV$^-$ between its ground $^3A_2$ and excited $^3E$ levels at a rate $\sigma_\mathrm{abs} I$, where $\sigma_\mathrm{12}$ is the NV$^-$ optical absorption cross-section and $I$ is the excitation intensity. NV$^-$ may then radiatively decay directly back to the ground state or non-radiatively decay via the intermediate $^1A_1$ and $^1E$ levels. Photoconversion to NV$^0$ occurs when an additional photon is absorbed when the center is in the excited $^3E$ level.\cite{waldherr11} This additional photon ionizes the excess electron of NV$^-$ to the diamond conduction band, which is then trapped at another defect with a total effective rate $\sigma_\mathrm{ion} I$, where $\sigma_\mathrm{ion}$ is ionization cross-section. The center may be returned to NV$^-$ by slow electron trapping processes with the effective rate $\gamma_\mathrm{trap}$.

As the photoconversion process involves two photon absorption events, the photoconversion rate is quadratically dependent on excitation intensity $I$ for intensities below saturation.\cite{waldherr11} However, when the optical transition of NV$^-$ is saturated, the photoconversion rate becomes linearly dependent on excitation intensity.\cite{waldherr11} This saturation condition is represented by the reduction of the four-level rate equation model to an effective two-level model [depicted in figure \ref{fig:photoconversion}(b)], which provides a simple description of the photoconversion dynamics.\cite{aslam13} The equations of the reduced model are

\begin{eqnarray}
\dot{p}_-(t) & = & -\sigma_\mathrm{ion}Ip_-(t)+\gamma_\mathrm{trap}p_0(t) \nonumber \\
\dot{p}_0(t) & = & \sigma_\mathrm{ion}Ip_-(t)-\gamma_\mathrm{trap}p_0(t)
\end{eqnarray}

where $p_-(t)$ and $p_0(t)$ are the probabilities of the center being NV$^-$ and NV$^0$, respectively. Considering a red pulse of duration $\tau$, given $\sigma_\mathrm{ion} I\gg\gamma_\mathrm{trap}$, the charge state probabilities following the pulse are approximately 
\begin{eqnarray}
p_-(\tau) & \approx & p_-(0)e^{-\sigma_\mathrm{ion}I\tau} \nonumber \\
p_0(\tau) & \approx & 1-[1-p_0(0)]e^{-\sigma_\mathrm{ion}I\tau}
\end{eqnarray}
where $p_-(0)$ and $p_0(0)$ are the initial probabilities. Hence, the charge state probabilities are inversely related single exponential functions of the pulse duration.

\subsection{ODMR electrometry}

\subsubsection{Electron spin resonances}

\begin{figure}[hbtp]
\begin{center}
\mbox{
\subfigure[]{
\includegraphics[width=0.5\columnwidth] {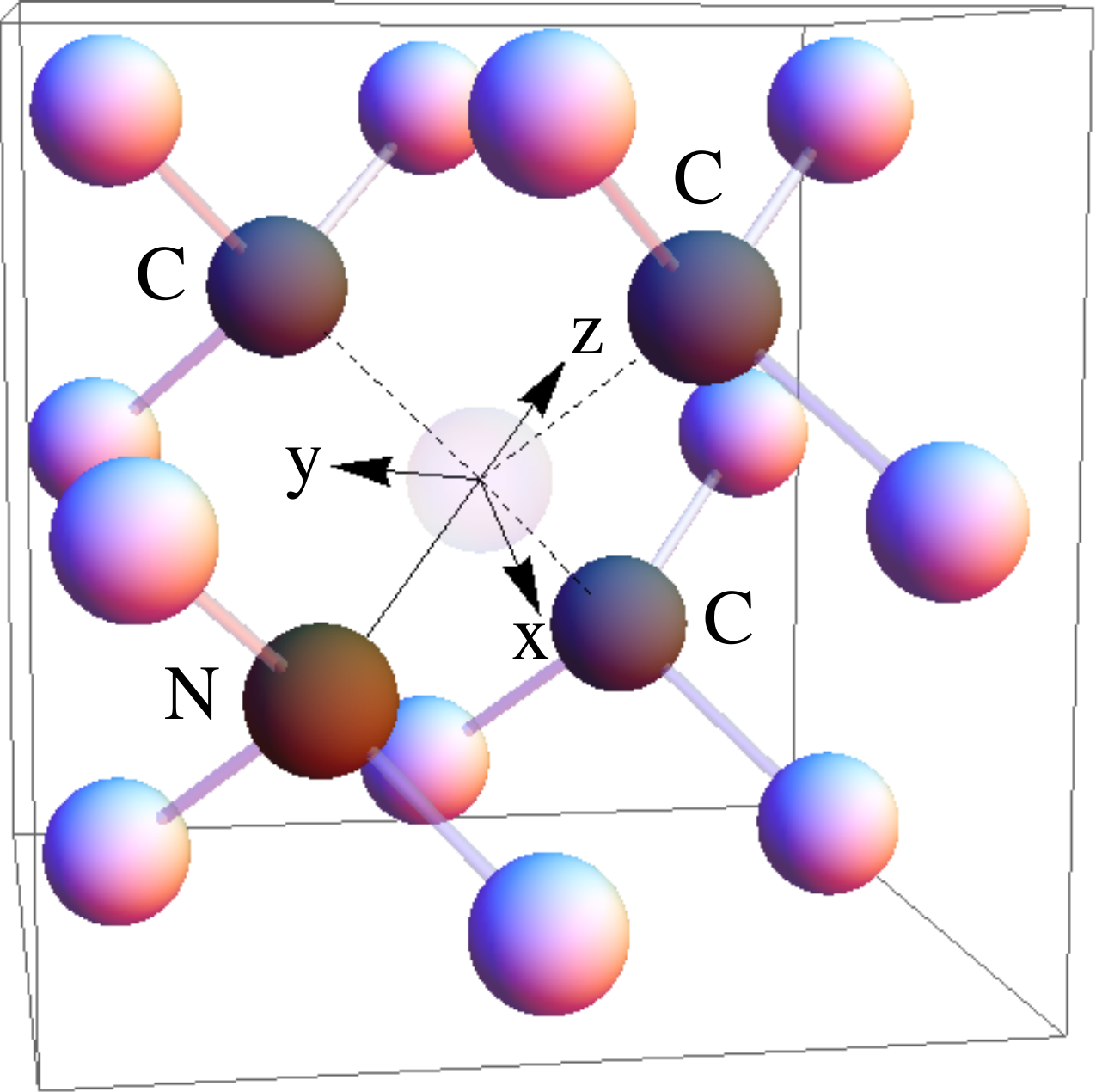}}
}
\mbox{
\subfigure[]{
\includegraphics[width=0.7\columnwidth] {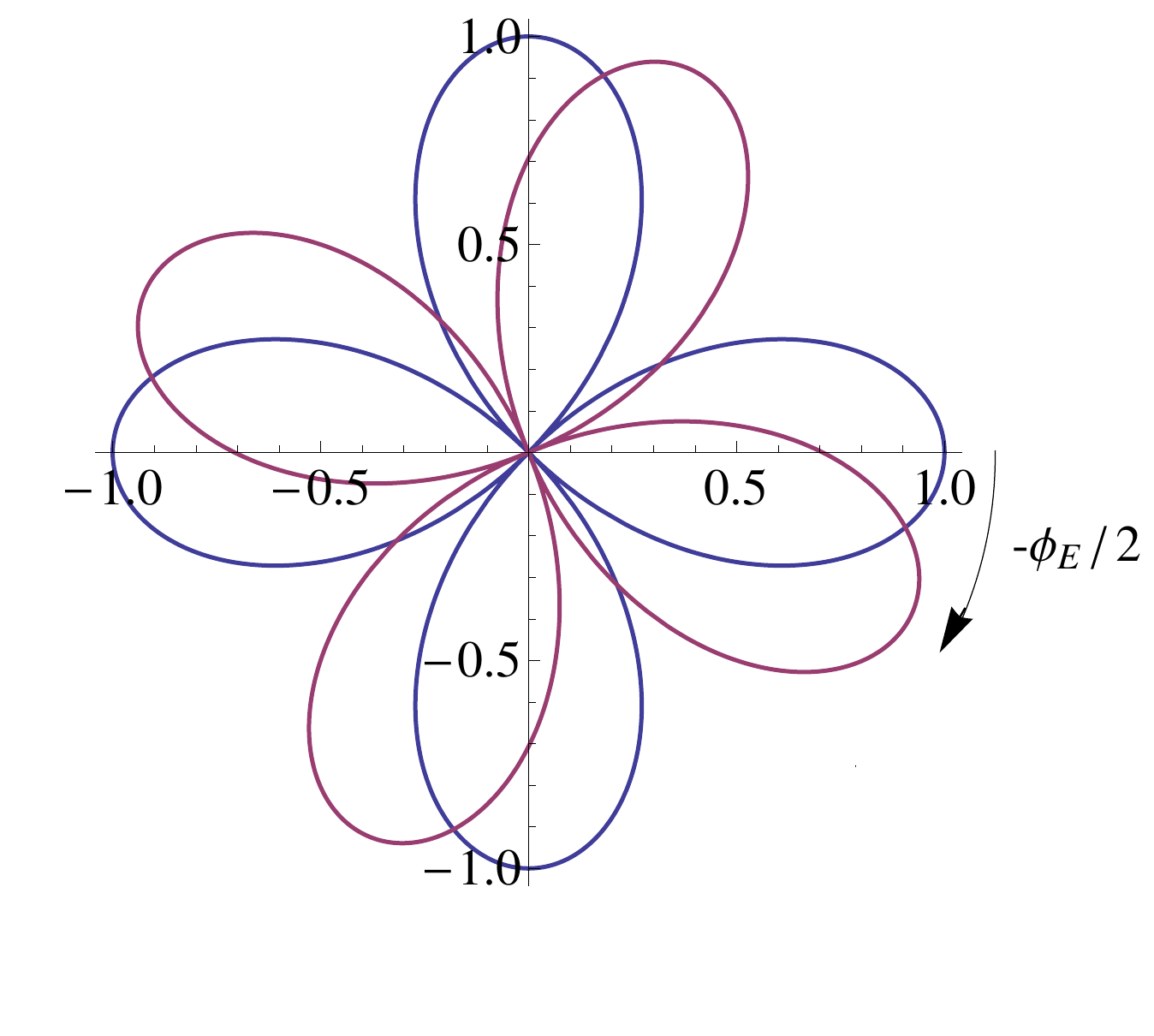}}
}
\caption{(a) Schematic of the NV center depicting the vacancy (transparent), the nearest neighbor carbon atoms (gray), the substitutional nitrogen atom (brown), the next-to-nearest carbon neighbors (white) and the adopted coordinate system ($z$ axis is aligned with the trigonal axis of the center and the $x$ axis is contained within the reflection plane defined by the nitrogen atom, vacancy and a nearest neighbor carbon atom). (b) Polar plot of the angular dependence $\cos(2\phi_B+\phi_E)$ of the spin resonances as the magnetic field is rotated in the transverse plane: Blue - $\phi_E=0$ and Red - $\phi_E=\pi/4$. The orientation of the `four-leaf' pattern is dependent on the orientation of the transverse electric field.}
\label{fig:spin1}
\end{center}
\end{figure}

The NV$^-$ ground state electron spin-Hamiltonian is \cite{doherty12}
\begin{eqnarray}
H&=&(D+ k_{\parallel} \Pi_z ) (S_z^2-2/3)+\gamma_e \vec{S}\cdot\vec{B}\nonumber \\
&&-k_{\perp}\Pi_x (S_x^2-S_y^2)+k_\perp \Pi_y (S_x S_y+S_y S_x )
\end{eqnarray}
where $\vec{S}$ are the $S=1$ dimensionless electron spin operators, $\gamma_e=g_e\mu_B/h$, $\mu_B$ is the Bohr magneton, $g_e\sim2.003$ is the electron g-factor,\cite{felton09} $h$ is the Planck constant, $\vec{B}$ is the magnetic field, $\vec{\Pi}=\vec{E}+\vec{\sigma}$ is the combined electric-strain field formed from the sum of the external electric field $\vec{E}$ and the effective electric field $\vec{\sigma}$ of local crystal strain, $k_\parallel=0.035(2)$ kHz m/V and $k_\perp=1.7(3)$ kHz m/V are the electric susceptibility parameters, \cite{vanoort90} and the spin coordinate system is defined such that the $z$ coordinate axis coincides with the center's trigonal symmetry axis and the $x$ axis is contained in one of the center's mirror planes (refer to figure \ref{fig:spin1}). Since transverse strain at NV~A can not be detected ($k_\perp\sqrt{\sigma_x^2+\sigma_y^2}<1$ kHz),\cite{dolde13} strain may be neglected and the above simplifies to
\begin{eqnarray}
H&=&(D+ k_{\parallel} E_z ) (S_z^2-2/3)+\gamma_e \vec{S}\cdot\vec{B}\nonumber \\
&&-k_{\perp}E_x (S_x^2-S_y^2)+k_\perp E_y (S_x S_y+S_y S_x )
\label{eq:spinhamiltonian}
\end{eqnarray}

Treating the electric and magnetic fields as perturbations to the zero-field splitting $D$, the approximate spin eigenstates (correct to first-order) are\cite{doherty12}
\begin{eqnarray}
\ket{z}^{(1)} & = &  \ket{z}^{(0)}-\frac{\gamma_eB_\perp}{\sqrt{2}D}\left(e^{-i\frac{\phi}{2}}s_{\frac{\theta}{2}}+e^{i\frac{\phi}{2}}c_{\frac{\theta}{2}}\right)\ket{-}^{(0)} \nonumber \\
&& -\frac{\gamma_eB_\perp}{\sqrt{2}D}\left(e^{-i\frac{\phi}{2}}c_{\frac{\theta}{2}}-e^{i\frac{\phi}{2}}s_{\frac{\theta}{2}}\right)\ket{+}^{(0)} \nonumber \\
\ket{-}^{(1)} & = & \ket{-}^{(0)}+\frac{\gamma_eB_\perp}{\sqrt{2}D}\left(e^{i\frac{\phi}{2}}s_{\frac{\theta}{2}}+e^{-i\frac{\phi}{2}}c_{\frac{\theta}{2}}\right)\ket{z}^{(0)} \nonumber \\
\ket{+}^{(1)} & = & \ket{+}^{(0)}+\frac{\gamma_eB_\perp}{\sqrt{2}D}\left(e^{i\frac{\phi}{2}}c_{\frac{\theta}{2}}-e^{-i\frac{\phi}{2}}s_{\frac{\theta}{2}}\right)\ket{z}^{(0)}
\end{eqnarray}
where
\begin{eqnarray}
&&\ket{z}^{(0)} = \ket{0} \nonumber \\
&&\ket{-}^{(0)} = e^{i\frac{\phi_E}{2}}\sin\frac{\theta}{2}\ket{1}+e^{-i\frac{\phi_E}{2}}\cos\frac{\theta}{2}\ket{-1} \nonumber \\
&&\ket{+}^{(0)} = e^{i\frac{\phi_E}{2}}\cos\frac{\theta}{2}\ket{1}-e^{-i\frac{\phi_E}{2}}\sin\frac{\theta}{2}\ket{-1}
\end{eqnarray}
$\tan\theta = k_\perp E_\perp/\gamma_eB_z$, $E_\perp=\sqrt{E_x^2+E_y^2}$, $B_\perp = \sqrt{B_x^2+B_y^2}$, $\phi = 2\phi_{B}+\phi_{E}$, $\tan\phi_{E}=E_y/E_x$, $\tan\phi_{B} = B_y/B_x$,  $c_{\frac{\theta}{2}} = \cos\frac{\theta}{2}$ and $s_{\frac{\theta}{2}} = \sin\frac{\theta}{2}$.
The approximate spin frequencies (correct to second-order) are\cite{doherty12}
\begin{eqnarray}
E_z^{(2)}/h & = & -\Lambda\nonumber \\
E_\pm^{(2)}/h & = & D+k_\parallel E_z + \Lambda \label{eq:spinfreq} \\ 
&&\pm{\cal R}
\sqrt{
1 -\left(\Lambda/{\cal R}\right) 2 \sin\theta\cos\phi+\left(\Lambda/{\cal R}\right)^2
}
\nonumber
\end{eqnarray}
where
\begin{eqnarray}
\Lambda & = & \frac{\gamma_e^2B_\perp^2}{2D} \nonumber \\
{\cal R}& = &\sqrt{k_\perp^2 E_\perp^2+\gamma_e^2B_z^2}.
\end{eqnarray}

Under the condition $\Lambda \gg {\cal R}$, the square root in $E_\pm^{(2)}$ can be expanded to first-order and the $\ket{z}\leftrightarrow\ket{\pm}$ spin resonance frequencies become linear in both axial and transverse electric field components
\begin{eqnarray}
f_\pm
&\approx& D + (2\pm1)\Lambda + k_\parallel E_z \mp{\cal R}\sin\theta\cos\phi \nonumber \\
 & = & f_\pm(0)+k_\parallel E_z\mp k_\perp E_\perp\cos(2\phi_B+\phi_E)
\label{eq:spinfreq_final}
\end{eqnarray}
where $f_\pm(0)= D+(2\pm1)\Lambda$.
The bare electric field effect can be expressed as $\Delta f_\pm = f_\pm - f_\pm(0) = k_\parallel E_z\mp k_\perp E_\perp\cos(2\phi_B+\phi_E)$. For small transverse electric field shift $k_\perp E_\perp$, the condition $\Lambda\gg\mathcal{R}$
can be realized by aligning the magnetic field in the transverse plane. Clearly, the larger $B_\perp$, the condition will be realized for a larger range of $B_z$, which was observed in this work. Note that the perturbative condition $\gamma_eB_\perp/D\ll1$ must still be satisfied for the above expression to be valid. 

The final term in
$f_\pm$
defines a coupled dependence of the spin resonance frequencies on the transverse orientations of the electric and magnetic fields. Polar plots of angular dependence of the spin resonances as the magnetic field is rotated in the transverse plane are depicted in figure \ref{fig:spin1}. The magnetic field rotation produces a characteristic `four-leaf' pattern. With knowledge of the transverse orientation of the magnetic field with respect to the trigonal structure of the NV center (which defines the coordinate system), the transverse orientation of the electric field $\phi_E$ can be determined up to a two-fold symmetry.\cite{efield}

\subsubsection{$^{15}\!$N Magnetic hyperfine interaction}

Since the observed electric field shifts are smaller than the $^{15}$N magnetic hyperfine interaction with the ground state electron spin, it is necessary to account for the hyperfine structure of the electron spin resonances. The hyperfine interaction is described by the addition of the following potential\cite{doherty12} to the electron spin-Hamiltonian (\ref{eq:spinhamiltonian})
\begin{eqnarray}
V_\mathrm{hf}=A_\parallel S_zI_z+A_\perp(S_xI_x+S_yI_y)
\end{eqnarray}
where $\vec{I}$ are the $I=1/2$ dimensionless nuclear spin operators, $A_\parallel=3.03(3)$ MHz and $A_\perp=3.65(3)$ MHz are the $^{15}$N hyperfine parameters.\cite{felton09} As described in the previous sub-section, in the presence of a transverse magnetic field, the $\ket{\pm}$ electron spin states are equal mixtures of the $m_s=\pm1$ spin projections. Consequently, there is no first-order magnetic hyperfine splitting of the $m_I=\pm1/2$ sub-levels of the $\ket{\pm}$ electron spin states. At second-order, the nuclear spin projections remain degenerate for $\ket{\pm}$, whilst they become equally mixed for the $\ket{z}$ electron spin state and split by $\delta f = 2A_\perp B_\perp/D$. Hence, the $\ket{z}\leftrightarrow\ket{\pm}$ electron spin resonances both split into two hyperfine resonances separated by $\delta f$.   If a small $B_z$ component is introduced to the magnetic field, the $\ket{\pm}$ electron spin states are no longer equal mixtures of the $m_s=\pm1$ spin projections and thus, the $\ket{\pm}$ states gain a first-order magnetic hyperfine splitting. Since the nuclear spin projections are still mixed in the $\ket{z}$ electron spin state, microwave transitions are allowed between each of the hyperfine levels of $\ket{z}$ and $\ket{\pm}$. Thus, in this case, the $\ket{z}\leftrightarrow\ket{\pm}$ electron spin resonances are both split into four hyperfine resonances. The hyperfine structure and its dependence on $B_z$ is summarized in figure 1 of the main article.

\subsubsection{Ramsey oscillations and the spin resonance spectrum}

The spin resonance spectrum was obtained by performing a Fast Fourier Transform (FFT) of the Ramsey oscillation that was observed by constructing a measurement ensemble through many iterations of the ODMR Ramsey pulse sequence and sweeping the Ramsey evolution time $\tau$. For a single spin resonance with frequency $f$, the observed Ramsey oscillation has the form\cite{doherty12}
\begin{eqnarray}
R(\tau) & \propto & e^{-\frac{\tau}{T_2}} \int_{-\infty}^\infty P(F)  \cos(2\pi F\tau) d F
\end{eqnarray}
where $F=f_\mathrm{mw}-f$, $f_\mathrm{mw}$ is the driving microwave frequency, $T_2$ is the homogeneous spin dephasing time and $P(F)$ is the statistical distribution of the resonance frequency (relative to $f_\mathrm{mw}$) within the inhomogeneous measurement ensemble. Performing a Fourier cosine transformation of $R(\tau)$, one obtains the spin resonance spectrum\cite{doherty12}
\begin{eqnarray}
R(\nu) & \propto & \int_{-\infty}^\infty P(F) L(\nu-F) d F
\end{eqnarray}
where $L(\nu-F)$ is the Lorentzian lineshape centered at $F$ and with width determined by $T_2$. It is clear that $P(F)$ leads to the inhomogeneous broadening of the spectral line. $P(F)$ is typically a single normal distribution related to a normally distributed source of noise that shifts the spin resonance. When the standard deviation of the distribution is much larger than $1/T_2$, the spectral line takes a Gaussian lineshape.

In our measurements, we tuned the driving microwave frequency $f_\mathrm{mw}$ to near the $\ket{z}\leftrightarrow\ket{-}$ electron spin resonance, which was distinct from the $\ket{z}\leftrightarrow\ket{+}$ electron spin resonance due to the transverse magnetic field. The spin resonance spectrum thus contained several spectral lines of the above form. Since there is a discrete electric field shift of the spin resonances of NV~A when NV~B switches from NV$^-$$\rightarrow$NV$^0$ and the excess electron is displaced to a distant location, the statistical distribution $P(F)$ of each spectral line of NV~A is composed of two normal distributions. Hence, after evaluating the statistical average, each spectral line becomes two lines with Gaussian lineshapes, whose intensities are related to the proportion of measurements that NV~B was in a given charge state.

\subsection{Fitting method}

The transverse electric field shift $k_\perp E_\perp$ at NV~A due to the change in charge at NV~B was determined by least squares fitting the observed spin resonances as functions of magnetic field using numerical solutions to the complete spin-Hamiltonian $H+V_\mathrm{hf}$. The fit contained several parameters: the magnitude of the fixed transverse magnetic field $B_y$, the orientation $\phi_E$ 
and magnitude $E_\perp$ of the transverse electric field, and the orientation $\theta_{\delta B}$ of the small magnetic field $\delta \vec{B}=\delta B \cos\theta_{\delta_B}\hat{x}+\delta B \sin\theta_{\delta_B}\hat{z}$ orthogonal to $B_y$ whose magnitude $\delta B$ was varied. The zero-field splitting parameter $D=2870.61(1)$ MHz of NV~A was independently determined using zero-field ODMR measurements. The $^{15}$N hyperfine parameters $A_\parallel=3.03(3)$ MHz and $A_\perp=3.65(3)$ MHz independently measured by Felton et al\cite{felton09} were also used.

As discussed in the main text, the axial electric field shift $k_\parallel E_z$ was defined to be zero in order to sufficiently constrain the fit of the transverse electric field shift $k_\perp E_\perp$. Fitting attempts where the axial electric field $E_z$ component was included as a free parameter did not converge due to the excessive flexibility of the set of fit parameters. The definition of $k_\parallel E_z\sim0$ is justified by the assessment that the $k_\parallel E_z$ would be too small to be detected, which is based upon the geometry of the two NV centers determined from super-resolution microscopy, the observed net electric field shift of $\sim66$ kHz, the orders of magnitude difference between $k_\parallel$ and $k_\perp$, and the assumption that the electric field at NV~A is orientated like an electric field of a point charge located at NV~B. This assumption corresponds to the physical picture that the observed electric field shift is due to the displacement of a charge (of arbitrary magnitude) at NV~B to a distant location, where the electric field of the charge could not be detected. Given the observed spectral linewidths, this distance was assessed to be $>40$ nm from NV~A.

Interpreting the observed electric field shift as being due to the displacement of a single electron at NV~B to a distant location, the transverse distance $r_\perp = 25\pm2$ nm to the electron from NV~A was calculated using the following formula that simply accounts for the relative permittivity of diamond ($\epsilon_r=5.7$)
\begin{eqnarray}
r_\perp^2 & = & 4\pi\epsilon_0\epsilon_r E_\perp / e^2
\end{eqnarray}
where  $e$ is the electron charge and $\epsilon_0$ is the vacuum permittivity. Since this distance is consistent with the distance obtained from super-resolution microscopy and there exists substantial evidence supporting the NV charge state assignments, we conclude that we have detected a single electron.